\documentclass[a4paper,fleqn,usenatbib]{mnras}

\usepackage{newtxtext,newtxmath}

\usepackage[T1]{fontenc}
\usepackage{ae,aecompl}

\usepackage{here}
\usepackage[pdftex]{graphicx}
\usepackage[pdftex]{color}
\usepackage{amsmath}	
\usepackage{amssymb}	

\makeatletter

\@addtoreset{equation}{section}
\makeatother
\title[Absorption line profiles from thermal winds]{Monte-Carlo simulations of the
detailed iron absorption line profiles from thermal winds in X-ray
binaries 
} 

\author[Tomaru et al.]
{Ryota Tomaru$^{2,1}$,\thanks{E-mail: tomaru@astro.isas.jaxa.jp}
Chris Done$^{3}$,
Hirokazu Odaka$^{4}$,
Shin Watanabe$^{1,2}$,
\newauthor
and Tadayuki Takahashi$^{5,1}$ \\
\\
$^{1}$Institute of Space and Astronautical Science , Japan Aerospace Exploration Agency , 3-1-1 Yoshinodai, Chuo, Sagamihara, Kanagawa, Japan\\
$^{2}$Department of Physics, University of Tokyo, 7-3-1 Hongo, Bunkyo, Tokyo, Japan\\
$^{3}$Department of Physics, University of Durham, South Road, Durham DH1 3LE, UK\\
$^{4}$High Energy Astrophysics Labolatory, Nishina Center for Accelerator-Based Science, RIKEN, 2-1 Hirosawa, Wako, Saitama, Japan\\
$^{5}$Kavli Institute for the Physics and Mathematics of the Universe, The University of Tokyo, 5-1-5 Kashiwanoha, Kashiwa, Chiba, Japan
}

\date{Accepted 2018 February 5. Received 2018 February 5; in original form 2017 December 27}

\pubyear{2018}

\begin{document}
\label{firstpage}
\pagerange{\pageref{firstpage}--\pageref{lastpage}}
\maketitle

\begin{abstract}

  Blue--shifted absorption lines from highly ionised iron are seen in some high inclination X-ray binary systems, indicating the presence of an equatorial disc wind. This launch mechanism is under debate, but thermal driving should be ubiquitous. X-ray irradiation from the central source heats disc surface, forming a wind from the outer disc where the local escape velocity is lower than the sound speed. The mass loss rate from each part of the disc is determined by the luminosity and spectral shape of the central source. We use these together with an assumed density and velocity structure of the wind to predict the column density and ionisation state, then combine this with a Monte-Carlo radiation transfer to predict the detailed shape of the absorption (and emission) line profiles. We test this on the persistent wind seen in the bright neutron star binary GX 13+1, with luminosity $L/L_{\mathrm{Edd}}\sim0.5$. We approximately include the effect of radiation pressure because of high luminosity, and compute line features. We compare these to the highest resolution data, the Chandra third order grating spectra, which we show here for the first time. This is the first physical model for the wind in this system, and it succeeds in reproducing many of the features seen in the data, showing that the wind in GX13+1 is most likely a thermal-radiation driven wind. This approach, combined with better streamline structures derived from full radiation hydrodynamic simulations, will allow future calorimeter data to explore the detail wind structure. 
\end{abstract}



\section{Introduction}

Low Mass X-ray Binaries (LMXBs) are systems where a companion star
overflows its Roche lobe, so that material spirals down towards a
compact object which can be either a black hole or neutron star. The
observed X-ray emission is powered by the enourmous gravitational
potential energy released by this material as it falls inwards,
lighting up the regions of intense space--time curvature and giving
observational tests of strong gravity.

This inflow also powers outflows. Accretion disc winds are seen via
blue--shifted absorption lines from highly ionised material in high
inclination LMXBs (see e.g. the reviews by \citealt{Ponti2012,
  DiazTrigo2016}) but the driving mechanism of these winds is not well
understood. The potential candidates are acceleration of gas by the
Lorentz force from magnetic fields threading discs (magnetic driving:
\citealt{Blandford1982, Fukumura2014}), radiation pressure on the
electrons (continuum driving: \citealt{Proga2002, Hashizume2015}) and
thermal expansion of the hot disc atmosphere heated by the central
X-ray source, which makes a wind at radii which are large enough for
the sound speed to exceed the local escape velocity (thermal driving:
\citealt{Begelman1983, Woods1996}, hereafter W96).

Recent work has focussed on magnetic driving, primarily because of a
single observation of dramatic wind absorption seen from the black
hole GRO J1655-40 at low luminosity, far below the Eddington limit
where continuum driving becomes important, and with a derived launch
radius which is far too small for thermal driving
\citep{Miller2006, Luketic2010, Higginbottom2015}. Magnetic wind
models can fit this spectrum \citep{Fukumura2017}, but then it is
difficult to understand why such high column and (relatively) low
ionisation winds are not seen from other observations of this object or any other high inclination systems with
similar luminosities and spectra.  Instead, the unique properties of
this wind could potentially be explained if the outflow has become
optically thick along the line of sight. This would suppress the observed flux
from an intrinsically super--Eddington source \citep{Shidatsu2016, Neilsen2016}.
Alternatively, this singular wind may be a transient phenomena,
not representative of the somewhat lower column/higher ionisation
winds which are normally seen. 

What then powers the more typical winds? Thermal driving qualitatively
fits the observed properties, as these winds are preferentially seen
in systems with larger discs \citep{DiazTrigo2016}.  \if{\bf\fi X-rays from
near the compact objects (the inner disc emission and any corona
and/or boundary layer) irradiate the outer disc, and the balance between
Compton heating and cooling heats the surface to the Compton
temperature, which is a luminosity weighted mean energy given by
$T_{\mathrm{IC}} = \frac{1}{4}\int E L(E)dE/\int L(E) dE$
(\citealt{Begelman1983, Done2018}, hereafter D18).\if}\fi
This temperature is constant with radius, as it
depends only on the spectrum of the radiation from the central region, but gravity decreases
with distance from the source. For a large enough disc, the 
isothermal sound speed from this Compton temperature is bigger than the escape
velocity, so the heated material makes a transition from forming a bound atmosphere, to an outflowing wind. 
This defines the Compton Radius
($R_{\mathrm{IC}}=GMm_p/kT_{\mathrm{IC}}= (6.4\times 10^4/T_{\mathrm{IC,8}}) R_g$, where
$R_g=GM/c^2$ and $T_{\mathrm{IC,8}}=T_{\mathrm{IC}}/10^8~K$) as the typical launch radius
for a thermal wind (\citealt{Begelman1983}, D18).

These thermal wind models give an analytic solution for the mass loss
rate from the disc (\citealt{Begelman1983}, W96). However, to calculate
the observables such as column density and ionisation structure
requires an understanding of the velocity as a function of
two-dimensional position (or equivalently, velocity as a function of
length along a streamline, together with the shape of the
streamlines). D18 assume a very simple 2-dimensional density and
velocity structure along radial streamlines, and show that this can
match the column seen in the full hydrodynamic simulation of W96. This
composite model was applied to multiple spectra of H1743--322, matching
the wind seen in its disk dominated state, and predicting the
disappearance of these wind absorption lines in a bright low/hard
state, as observed (Shidatsu \& Done 2017, in prep).  This wind
disappearance does not occur from simply the change in photo-ionisation
state from the changing illumination \citep{Miller2012}. The key
difference is that thermal winds respond to changing illumination by
changing their launch radius, density and velocity as well as
responding to the changing photo-ionising spectrum (Shidatsu \& Done
2017, in prep).

Here we explore the thermal wind predictions in more detail, using a
3--dimensional Monte-Carlo simulation code MONACO \citep{Odaka2011} to calculate
the radiation transport through the material so as to predict the
resulting emission and absorption line profiles.  We calculate these
for the very simple constant velocity structure assumed in D18, and
then extend this  to consider an accelerating wind along biconical
streamlines, as a more appropriate disk wind geometry
\citep{Waters2012}.  We show results from both these geometries for
the specific case of $L/L_{\mathrm{Edd}}=0.3, T_{\mathrm{IC,8}}=0.13, R_{out}=5R_{\mathrm{IC}}$
to compare with W96.  

\if{\bf\fi We apply the biconical wind model to the bright neutron star binary
system GX13+1. This system is unique amongst all the black hole and
neutron star binaries in showing persistently strong blueshifted
absorption features in its spectrum  \citep{Ueda2004,D'Ai2014}.
The X-ray continuum is rather
stable, with $ T_{\mathrm{IC,8}}\sim 0.13$, making it a good match to
the simulations, but it has slightly higher luminosity at 
$L/L_{\mathrm{Edd}}=0.5$. This means that  radiation pressure should become
important, decreasing the radius from which the wind can be launched
and increasing its column density. \if}\fi
We compare results from a hybrid
thermal/radiative wind (calculated using the approximate radiation
pressure correction from D18) to the detailed absorption line profiles
seen in the third order Chandra grating data from this source.  This
is the first quantatiative, physical model for the wind, and the first
exploration of the highest spectral resolution data from this
source. \if{\bf\fi While magnetic driving models are always possible, \if}\fi
our results match the majority of the observed features,
showing that the wind properites are broadly consistent with hybrid
thermal-radiative driving.

\section{Radiative Transfer Code}

We use the Monte-Carlo simulation code MONACO \citep{Odaka2011} to
calculate radiative transfer through the wind.  
MONACO uses its original physics implementation of photon interactions \citep{Watanabe2006, Odaka2011} while it employs the Geant4 toolkit library \citep{Agostinelli2003} for photon tracking in an arbitrary 3--dimensional geometry.

We consider an
azimuthally symmetric density and velocity field, then use the XSTAR
photo-ionization code \citep{Kallman2001} to calculate the
equilibrium population of ions from each element assuming
one-dimensional radiation transfer from the central source. We 
grid in radial distance, $r$, and $\theta$ and assume the illuminating
flux is the transmitted part of the central spectrum along the line of
sight to the element. This gives an ionisation parameter
\begin{equation}
\xi (r,\theta) = \frac{L \exp(-\tau_{abs})}{n (r,\theta) r^2}
\end{equation}
where $\tau_{abs}$ is the optical depth to absorption (but does not
include electron scattering), and $n(r,\theta)$ is the density as a function of position. 

We then use MONACO to track photons through this ionisation structure,
including their interaction with this material. Photons interacting
with ions can be absorbed in photo-ionisation or photo-excitation, and photons generated via recombination and atomic
deexcitation are tracked. \if{\bf\fi Doppler shifts of the absorption cross-sections
from the velocity structure of the material
are included\if}\fi, as is the Compton energy change on interaction with free
electrons. Ideally, this calculated radiation field should then be
used as input to XSTAR, the ion populations recalculated,  and the
process should be iterated until convergence. However, for our simulations here
the wind is mostly optically thin, so we do not include this
self-consistent iteration. 
This method of radiation transfer calculation is based on the \citet{Hagino2015}, 
but the geometry and the velocity/density structure we use is reflected on the 
thermally driven winds in LMXBs  whereas \citet{Hagino2015} is focussed on 
the UV-line driven winds in Active Galactic Nucleus (AGN).
Also, as thermal winds are typically
highly ionised, we consider only H-like and He-like ions of
Fe and calculate the spectrum only over a restricted energy band of
6.5--7.2 keV. Table.\ref{table:lines} details the transitions used. 

\begin{table}
\begin{tabular}{|c|c|c|}\hline
Line ID & Energy [keV] & oscillator strength  \\  \hline
Fe {\scriptsize XXV} He$\alpha  ~\mathrm{y}$ & 6.668  & $6.57 \times 10^{-2}$  \\ 
Fe {\scriptsize XXV} He$\alpha~ \mathrm{w}$ & 6.700 & $7.26 \times 10^{-1} $\\
Fe {\scriptsize XXVI} Ly$\alpha_2$ &6.952& $1.36\times 10^{-1}$ \\
Fe {\scriptsize XXVI} Ly$\alpha_1$ &6.973& $2.73\times 10^{-1}$ \\ \hline
\end{tabular}
\caption{Detailed parameters for each line included in these MONACO simulations. Note that we list only lines which have larger oscillator strengths than $10^{-3}$. }
\label{table:lines}
\end{table}

\section{Radial streamlines: D18}

\subsection{Geometry and Parameters}

We first consider the radial streamline wind model of D18. This
calculates the analytic mass loss rate per unit area, $\dot{m}(R)$
where $R$ denotes distance along the disc plane.  Integrating over the
whole disc gives the total mass loss rate in the wind, $\dot{M}$.
This is assumed to flow along radial (centered at the origin)
streamlines from a launch radius which is $0.2R_{\mathrm{IC}}$ for high
$L/L_{\mathrm{Edd}}$, with constant velocity set at the mass loss weighted
average escape velocity. The mass loss rate along each radial
streamline is weighted with angle such that 
$\dot{M}(\theta)\propto \dot{M}(1-\cos\theta)$, and then mass
conservation gives 
$n(r,\theta) \propto ( 1-\cos\theta)/r^2$.  D18 show that
these assumptions lead to a total column density through the structure
which matches to within a factor 2 of that in the hydrodynamic
simulations of W96 (see also Section 4).

We put this structure into MONACO for $L=0.3
L_{\mathrm{Edd}}$ with $T_{\mathrm{IC}}=1.3\times 10^7$~K and $R_{\mathrm{out}}=5R_{\mathrm{IC}}$ (mass
loss rate $\dot{M}_w = 2.0 \times 10^{19}  \mathrm{g~s^{-1}}$ for a $10M_\odot$ black hole which means 
the ratio of mass loss rate to mass accretion rate $\dot{M}_w/\dot{M}_a = 3.9$, where the $\dot{M}_a = L/(0.1c^2)$,
launch radius of $0.2R_{\mathrm{IC}}\approx 10^5R_g$ and 
weighted average $v_{\mathrm{out}}=420\mathrm{km/s}$). We include
turbulence, assuming $v_{\mathrm{turb}}=v_{\mathrm{out}}$, and 
calculate the rotation velocity
along each stream line assuming angular momentum conservation (see
Appendix A).

\begin{figure}
\includegraphics[width=0.9\hsize]{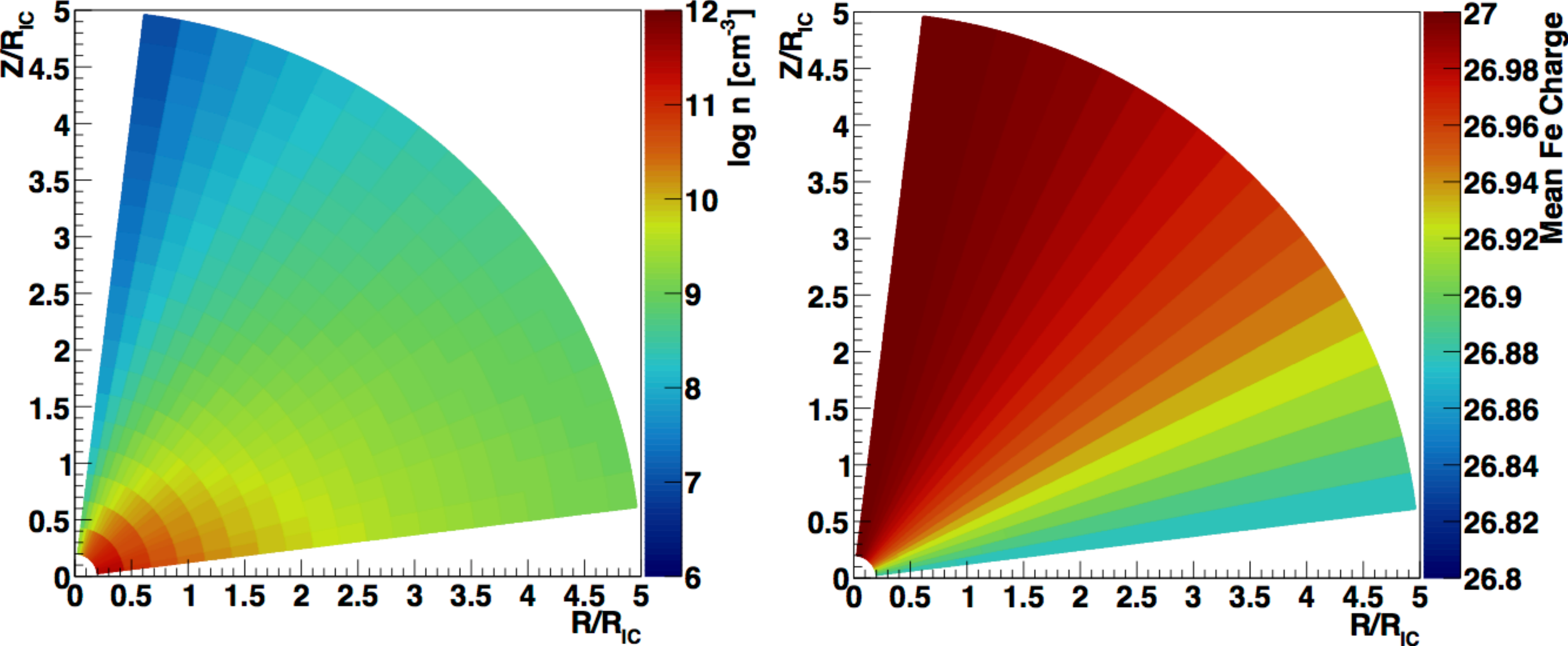}
\caption{Distribution of density (left) and mean Fe ionization state
  (right) for the radial streamline model}
\label{fig: monaco input s}
\end{figure}

We make a grid which follows the symmetry of the assumed structure,
i.e. centred on the origin, with 20 linearly spaced spherical shells
from $0.2-5R_{\mathrm{IC}}$, and 20 angles, linearly spaced in $\theta$ from
$7-83^\circ$ (see below). This density structure is shown in the left
panel of Fig.~\ref{fig: monaco input s}, while the right panel shows
the mean Fe ion state obtained from the XSTAR calculation.  This is
constant along each streamline because the ionisation parameter
$\xi=L/(n r^2)$, and the constant velocity radial streamlines mean
that density decreases as $1/r^2$.  Fe is almost completely ionised
over the whole grid, with a small fraction of hydrogen-like iron
remaining only for high inclination streamlines.

\subsection{MONACO output}

\begin{figure*}
 \includegraphics[width=\hsize]{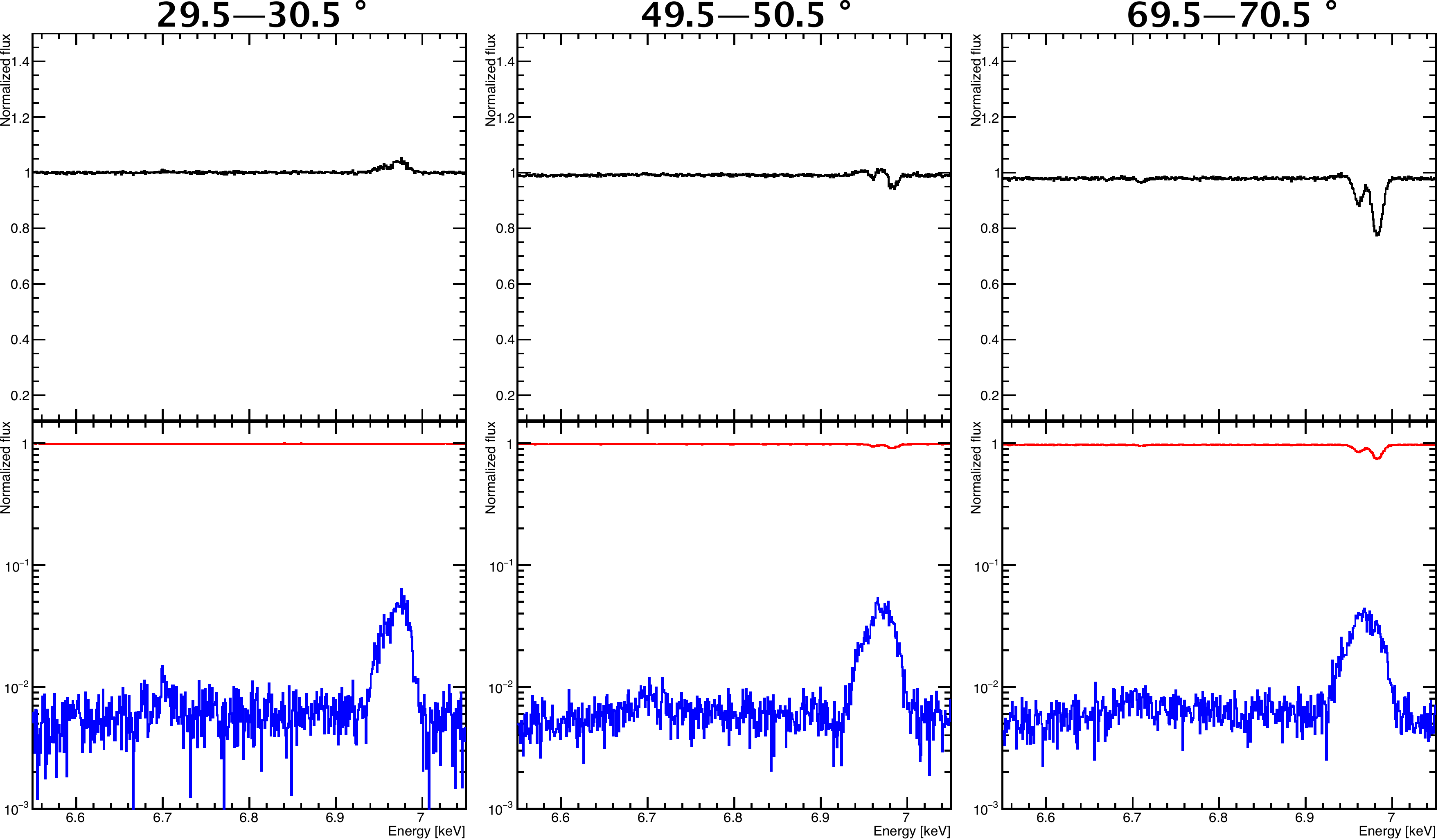}
\caption{ Spectra computed for the radial streamline model with 1 eV resolution for different lines of sight. Each panel shows the spectrum in a different inclination angle bin (the angular bin sizes are indicated the top of each panel).  The total spectrum is shown in black (top), the spectrum direct photons in red and scattered/reprocessed spectrum is blue (bottom). Note that the vertical axis is plotted linearly in the top panels but logarithmically in the bottom panels.   Lines are Fe {\scriptsize XXV} ~(6.668 keV for $\mathrm{He\alpha} ~y$ and 6.700 keV for $\mathrm{He\alpha} ~w $ ) and Fe {\scriptsize XXVI}~ (6.952 keV for $\mathrm{Ly\alpha_2}  $ and 6.973 keV for $\mathrm{Ly\alpha_1})$.
The equatorial density structure of the wind means that the absorption is much stronger at high inclination angles. The 
emission is more isotropic, so it can clearly be seen at low inclination angles, but is absorbed by the wind at high 
inclinations. }
\label{fig: spectra s}
  \end{figure*}

\begin{figure}
  \includegraphics[width=0.9\hsize]{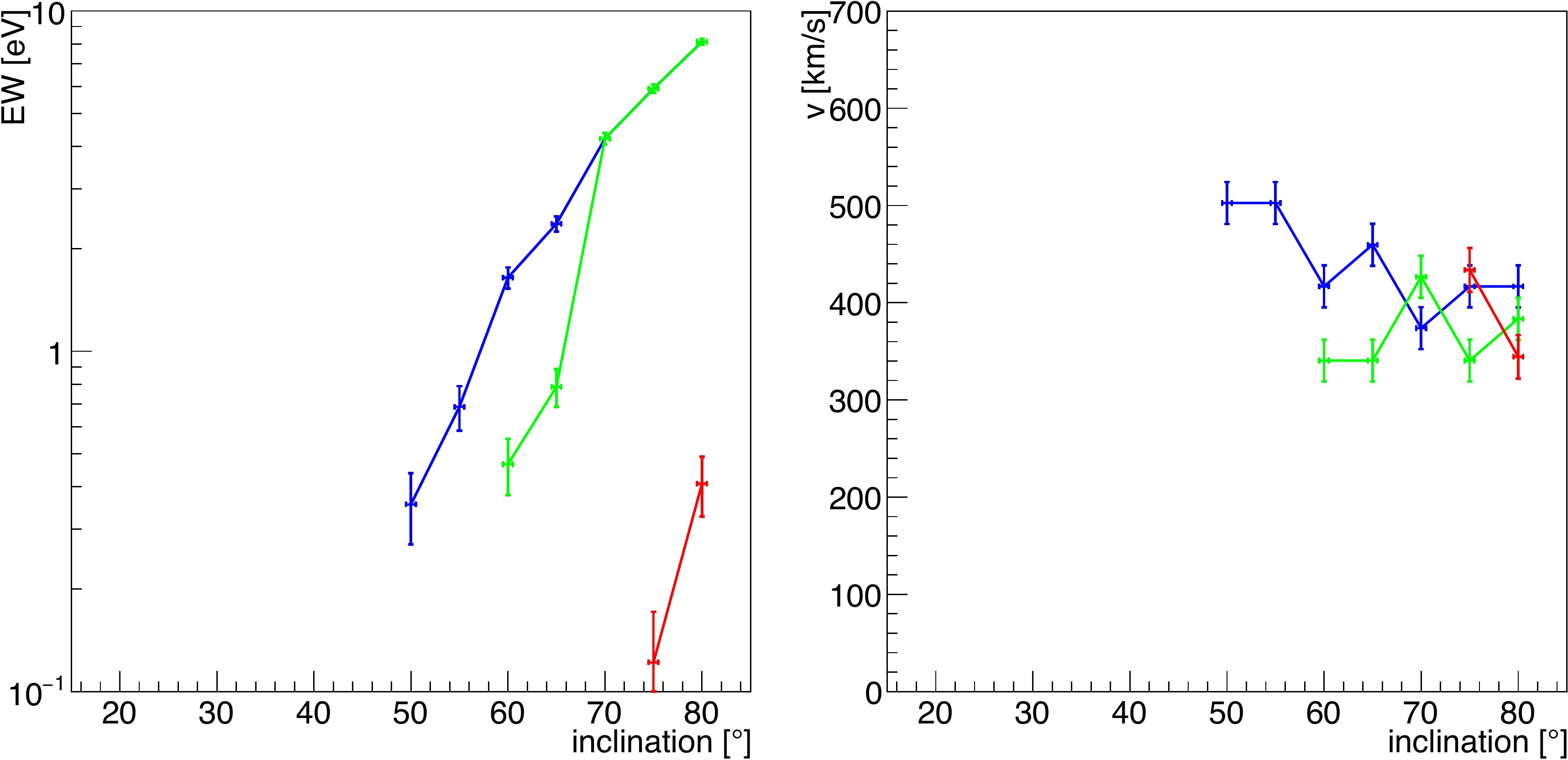}
  \caption{Left panel: EW of the absorption lines as a function of
    inclination angle, Fe {\scriptsize XXV} ($\mathrm{He \alpha ~w}$,
    red) and Fe {\scriptsize XXVI} ($\mathrm{Ly \alpha_2}$, green and
    $\mathrm{Ly\alpha_1}$, blue). The EW of all absorption lines
    increases strongly at higher inclination, showing the assumed
    equatorial disc wind geometry. The Doppler wings (with width set
    by turbulent velocity) of the two H-like absorption lines start to
    merge for inclinations above $70^\circ$ so above this we show the
    total EW of the two lines. Right panel: the blue shifted
    absorption line velocity for each ion species. This clearly shows
    the assumed constant velocity structure of the radial streamline.}
 \label{fig: ew s}
  \end{figure}

Fig.~\ref{fig: spectra s} shows the resulting spectra at three
different inclination angles. \if{\bf\fi These show that the emission lines
  are always similarly weak, and that the electron scattered continuum
  flux makes only a $\sim 0.5$\% contribution to the total flux, but
  that the absorption lines strongly increase at higher inclination
  angles. We calculate the equivalent width (EW) of each emission and
  absorption line by fitting the continuum outside the emission and
  absorption regions with an arbitrary function ($F(E) = aE^{bE+c}$
  \citealt{Odaka2016}). The EW of each emission and absorption line is
  then measured by numerical integration of the difference between the
  model and the simulation data. The left panel of Fig.~\ref{fig: ew
    s} shows the EW of the He-like (red) and H-like Ly$\alpha_2$
  (green) and Ly$\alpha_1$ (blue) absorption lines. The corresponding
  emission lines always have EW lower than $0.1$~eV so are not seen on
  this plot.  The strong
  increase of the absorption line EW with inclination clearly
  shows that the wind is equatorial (by construction from the $1-\cos
  \theta$ density dependence and constant velocity assumptions). At
  inclinations above $70^\circ$, the Doppler wings of the K$\alpha_1$
  and K$\alpha_2$ absorption lines merge together due to the turbulent
  velocities, so Fig.~\ref{fig: ew s} shows only a single EW for this blend.
\if}\fi

Fig.~\ref{fig: ew s} (right) shows the outflow velocity, as measured
from the energy of the deepest absorption lines (with error set by the
resolution of the simulation to $\pm 0.5$~eV). These velocities are
constant within 25 \% as a function of inclination, again by construction due to
the assumption of constant radial velocity along radial
streamlines. 

\section{Diverging wind}

In section 3, we considered a wind model with constant velocity along
radial streamlines.  However, the expected thermal wind geometry is
instead much more like an accelerating, diverging biconical wind \citet{Waters2012}.  Full
streamline structures which give the density and velocity of the wind
at all points can only be found by hydrodynamic calculations (but see
\citealt{Clarke2016} for some analytic approximations). Since modern
calculations only exist for the singular case of GRO J1655-40, we
follow D18 and use the W96 simulation results. W96 does not give full density/velocity structures, but do 
give  total column density through the wind at three different luminosities.
We use these to
match to our assumed streamline and velocity structure, which is the
standard biconical diverging disc wind used in a variety of systems
including cataclysmic variables \citep{Knigge1995,Long2002} and Active
Galaxies \citep{Sim2010,Hagino2015}.

\subsection{Geometry and Parameters} 

\begin{figure}
 \includegraphics[width=0.9\hsize]{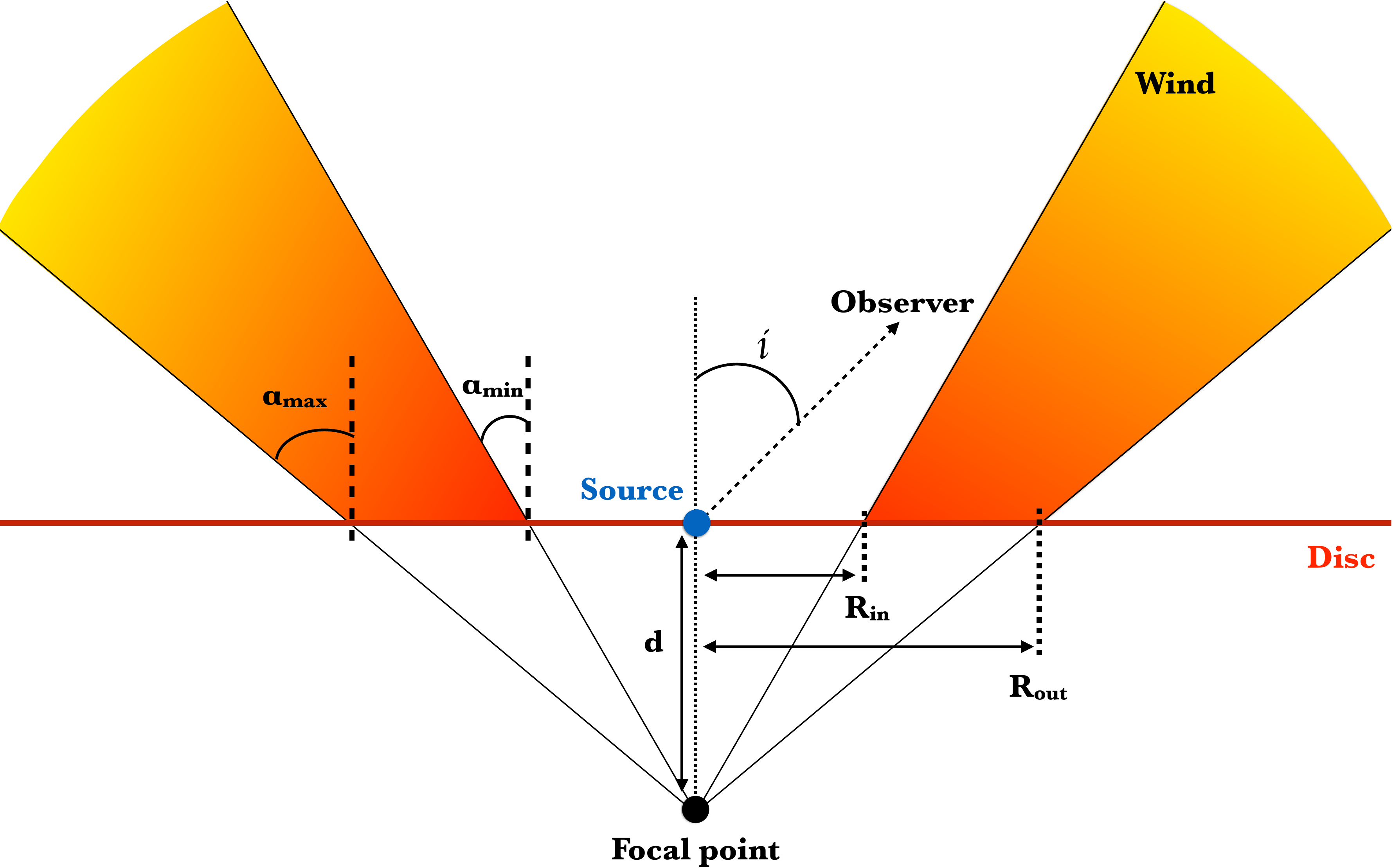}
 \caption{The geometry of the diverging biconical wind model.}
 \label{fig:diverging}
\end{figure}

The geometry can be defined by 3 parameters (Fig.~\ref{fig:diverging}).
\begin{enumerate}
\item $R_{\mathrm{in}}=0.1R_{\mathrm{IC}}$, the distance from the source to the inner edge of the wind
\item $R_{\mathrm{out}}$, the distance from the source to outer edge of the wind
\item $\alpha_{\mathrm{min}}$, the angle from z axis to the inner edge of the wind
\end{enumerate}
The disc wind is fan-shaped, with a focal point offset down from
the centre by a distance $d= 0.1R_{\mathrm{IC}}/\tan\alpha_{\mathrm{min}}$ so that the
wind fills the angles from $\alpha_{\mathrm{max}}-\alpha_{\mathrm{min}}$ down to the
disc surface.  We use $R$ to denote distance along the disc
surface, and $r,\theta$ denote radial distance and polar angle from the
origin, as before.  $\alpha_{\mathrm{min}}$ (or equivalently $d$) is a free parameter,
which sets the wind geometry. 

Streamlines are assumed to be along lines of constant angle $\alpha$ (where
$\alpha_{\mathrm{min}}<\alpha<\alpha_{\mathrm{max}}$) from the 
focal point. 
Distance along a streamline which starts on the disc
at radius $R$ is $l(R)$ (see Appendix. \ref{sec:calculation model A}). Velocity along the
streamline is assumed to be of the form $v (r,\theta)=f_v
c_{ch}(r)\sqrt{\frac{l(r,\theta)}{R(r)}}$, i.e. this wind
accelerates with distance along the streamline, with a terminal velocity
which is related via a free parameter $f_v$ to the characteristic
sound speed $c_{ch}$, given by the balance between heating and cooling
in the time it takes the wind to reach a height $H\sim R$ (D18).  The
density structure is solved by the mass conservation continuity
equation along streamlines (see Appendix B).
We calculate the wind properties out to a distance which is twice
that of the focal point of the wind to $R_{\mathrm{out}}$.

\begin{figure}
 \includegraphics[width=0.9\hsize]{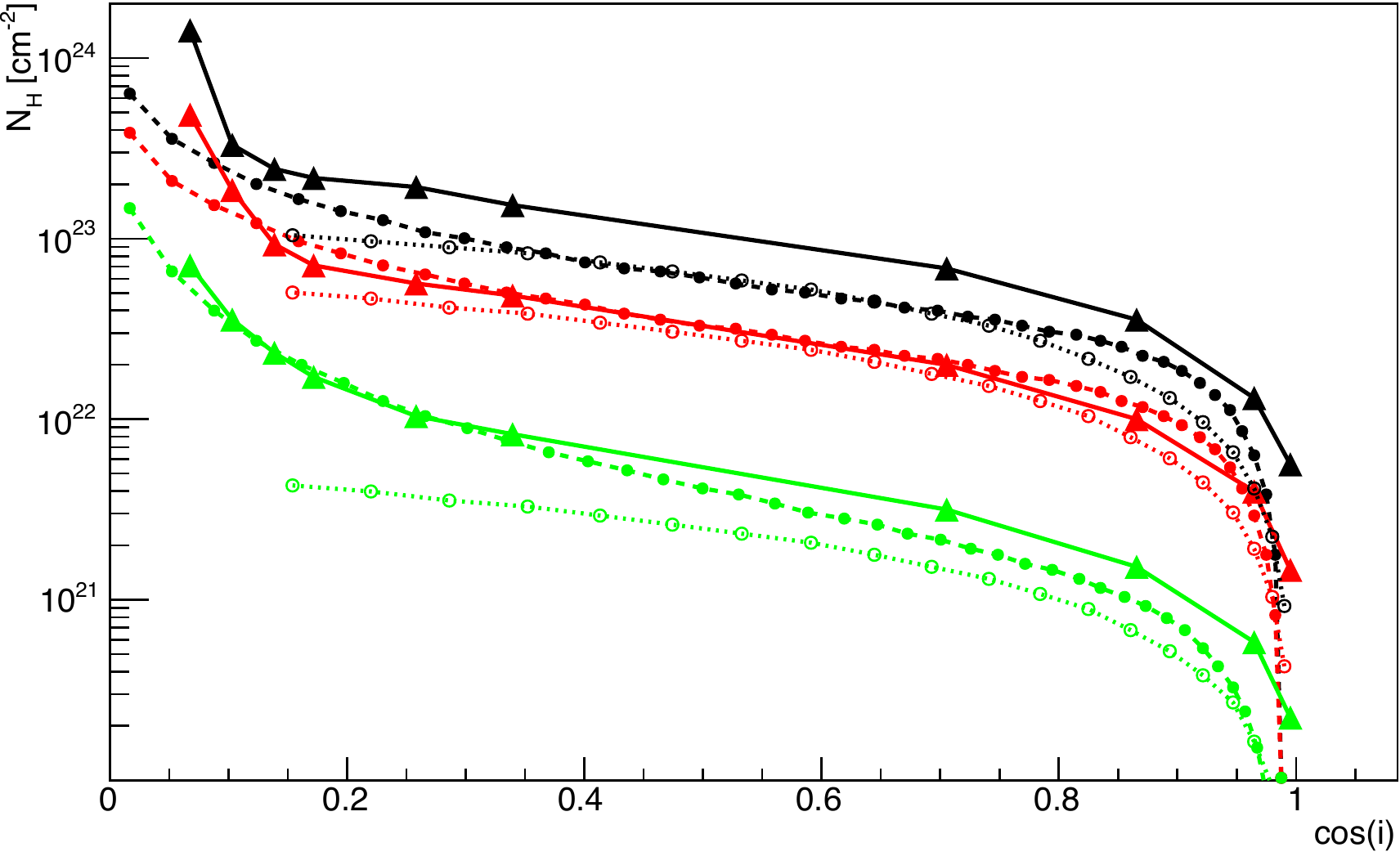}
 \caption{The solid lines show
column density as a function of the cosine of the inclination angle
through the wind resulting from the hydrodynamic simulations of W96
   for $L/L_{\mathrm{Edd}}=$ 0.01(green), 0.08 (red), 0.3 (black). The filled
   circles show that resulting from the diverging biconical wind
   (Section 4) while the open circles show the radial streamline model
   of D18 (Section 3).}
 \label{fig:column density}
\end{figure}

We set the free parameter values, $f_v$ and $\alpha_{\mathrm{min}}$, and
calculate the total column along each line of sight, $N_H(\theta)$, to
the central source for parameters matching to the three W96
simulations. These are $L/L_{\mathrm{Edd}} =0.3, 0.08$ with $R_{\mathrm{out}}=5R_{\mathrm{IC}}$
and $L/L_{\mathrm{Edd}}= 0.01$ with $R_{\mathrm{out}}=12R_{\mathrm{IC}}$. We adjust $f_v$ and
$\alpha_{\mathrm{min}}$ to minimise the difference between our model and
W96. We find $\alpha_{\mathrm{min}}=7^\circ$ and $f_v=0.25$ matches within a
factor 2 of the results from W96.  Fig.~\ref{fig:column density} shows
results with these parameters (filled circles), compared to the radial
wind model of Section 3 (open circles) as well as the W96 results
(solid line). This more physically realistic geometry and velocity 
gives a similarly good match to the simulations as the D18 radial wind.

\begin{figure}
\includegraphics[width=0.9\hsize]{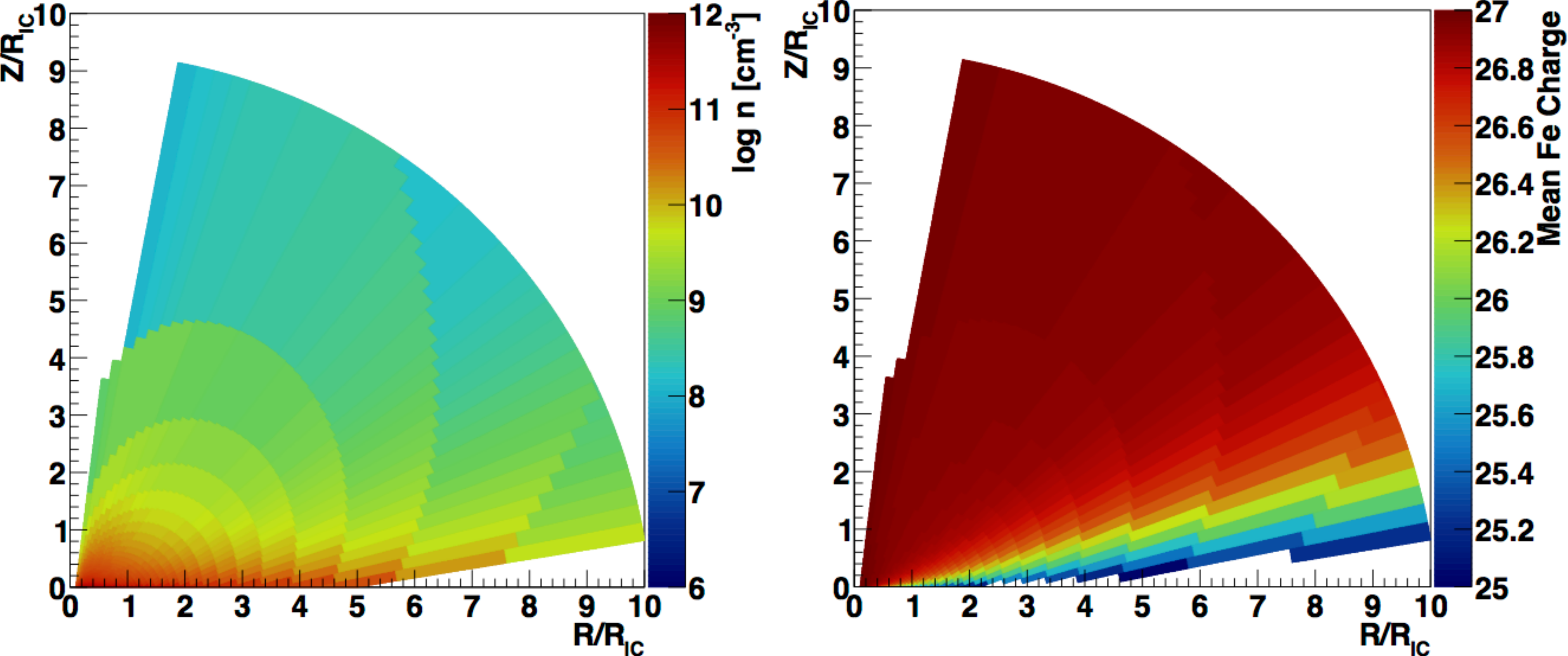}
\caption{Distribution of density (left) and Fe ionisation state (right)
for the diverging wind geometry. The accelerating flow gives
higher density material close to the disc compared to the constant
velocity outflow model in Fig.~\ref{fig: monaco input s}, giving 
lower mean ionisation state.}
\label{fig:xout d}
\end{figure}

The resulting density structure from this different geometry and
velocity are shown in the left panel of Fig.~\ref{fig:xout d} for
$L/L_{\mathrm{Edd}}=0.3$. Comparing this with the radial wind shows
that the density is higher closer to the disc, and lower further away
due to the material accelerating away from the disc rather than being
at constant velocity.  We run XSTAR as before, and the right panel of
Fig.~\ref{fig:xout d} shows that this leads to a lower mean ionisation
state of Fe than before, and this is no longer constant along the
radial sightline due to the different wind geometry.


\subsection{MONACO output}

\begin{figure*}
\includegraphics[width=\hsize]{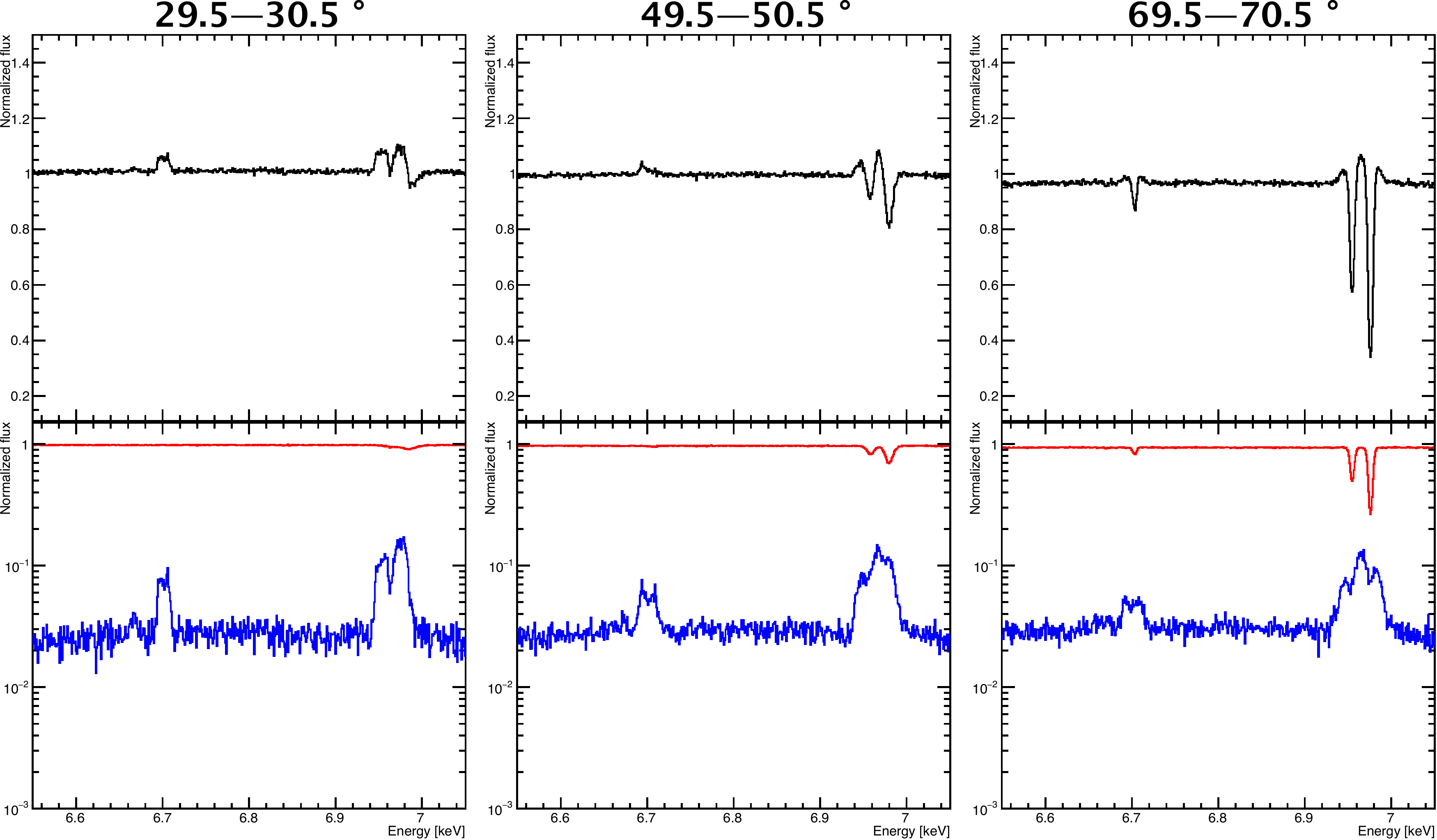}
\caption{As in Fig \ref{fig: spectra s}, but for the diverging biconical
wind geometry.}
\label{fig: spectra d}
\end{figure*}

\begin{figure}
\includegraphics[width=0.9\hsize]{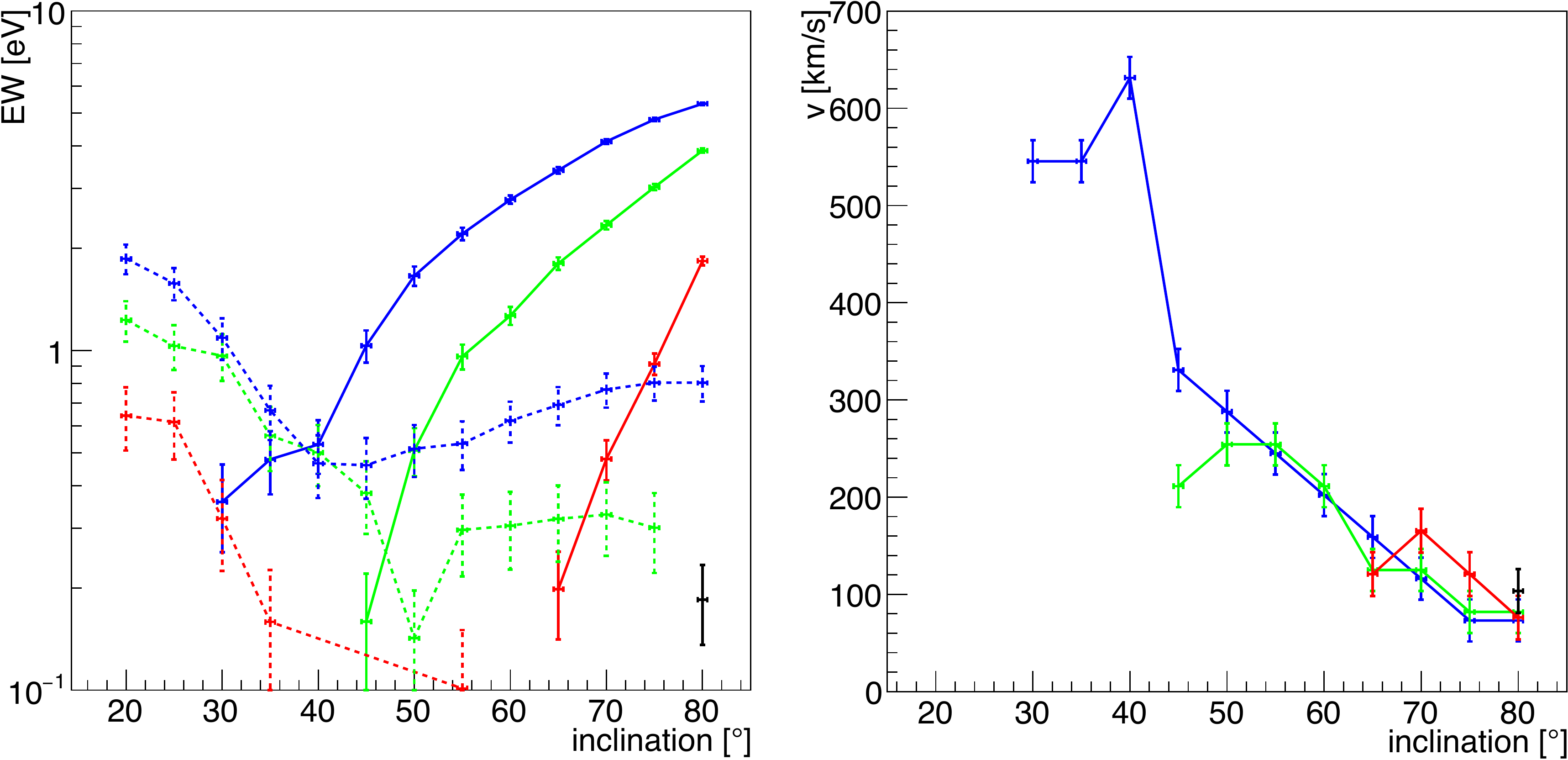}
\caption{As in Fig.~\ref{fig: ew s}, but for the diverging wind model.
The lower 
The lower ionisation state means that there is also a contribution
from the intercombination line of  Fe {\scriptsize XXV} $\mathrm{He \alpha ~y}$
(black) at the highest inclinations. }
\label{fig: ew d}
\end{figure}

We calculate the emission and absorption lines resulting from the
different wind structure (Fig.~\ref{fig: spectra d}).  The diverging
bipolar wind has 
higher density material closer to the source compared to the radial
wind geometry, so it subtends a 
larger solid angle to scattering. This means that there is 
more emission line contribution, as well as
a higher fraction of electron scattered continuum (around 2\%, see
the lower panel of Fig.~\ref{fig: spectra d}). The 
left panel of Fig.~\ref{fig: ew d} shows the emission line
EW (dotted lines) for each ion species (red: Fe XXV w,
green: Fe XXVI Ly$\alpha_2$, blue: Ly$\alpha_1$). These can now be of order
1eV for face on inclinations, decreasing at higher inclination as
they are significantly suppressed by line absorption. 

The corresponding absorption line EWs are shown as the solid lines
(compare to Fig.~\ref{fig: ew s}). The lower mean ionisation state
leads to more He-like Fe, so there is more of this ion seen in
absorption than in the radial streamline model.
These absorption lines increase as a function of inclination
angle as before, but now the Ly$\alpha_1$ and Ly$\alpha_2$ do not
merge together at the highest inclination angles due to the different
velocity structure (see right panel of Fig.~\ref{fig: ew d}). The
lines are formed preferentially in the higher density material
close to the disc. The assumed acceleration law means that the
typical velocities here are lower than in the constant velocity model,
as the material has only just begun to accelerate. Thus the turbulence
is also lower, so the Doppler width of the absorption lines is
smaller. This also means that the absorption line saturates to a
constant EW at lower column density, so the EW of the absorption lines
does not increase so strongly as before at the highest inclination angles. 

\section{Comparison with GX13+1}

We now use the more physically motivated diverging biconical wind
geometry to compare with observational data. An ideal source would be
one which is not too different from the parameters simulated in the
previous sections, as here we know the total column from W96 and know
that our assumed velocity/density matches to this.  Of the sources
listed in \citet{DiazTrigo2016}, the neutron star LMXB GX13+1 is the
source which has most similar $L/L_{\mathrm{Edd}}$ and
$T_{\mathrm{IC}}$ to that assumed here, and it also has the advantage
that it is a persistent source, with relatively constant luminosity
and spectral shape, and it shows similarly strong absorption lines in
multiple datasets.

\subsection{Observational data} 

\begin{figure}
 \includegraphics[width=0.9\hsize]{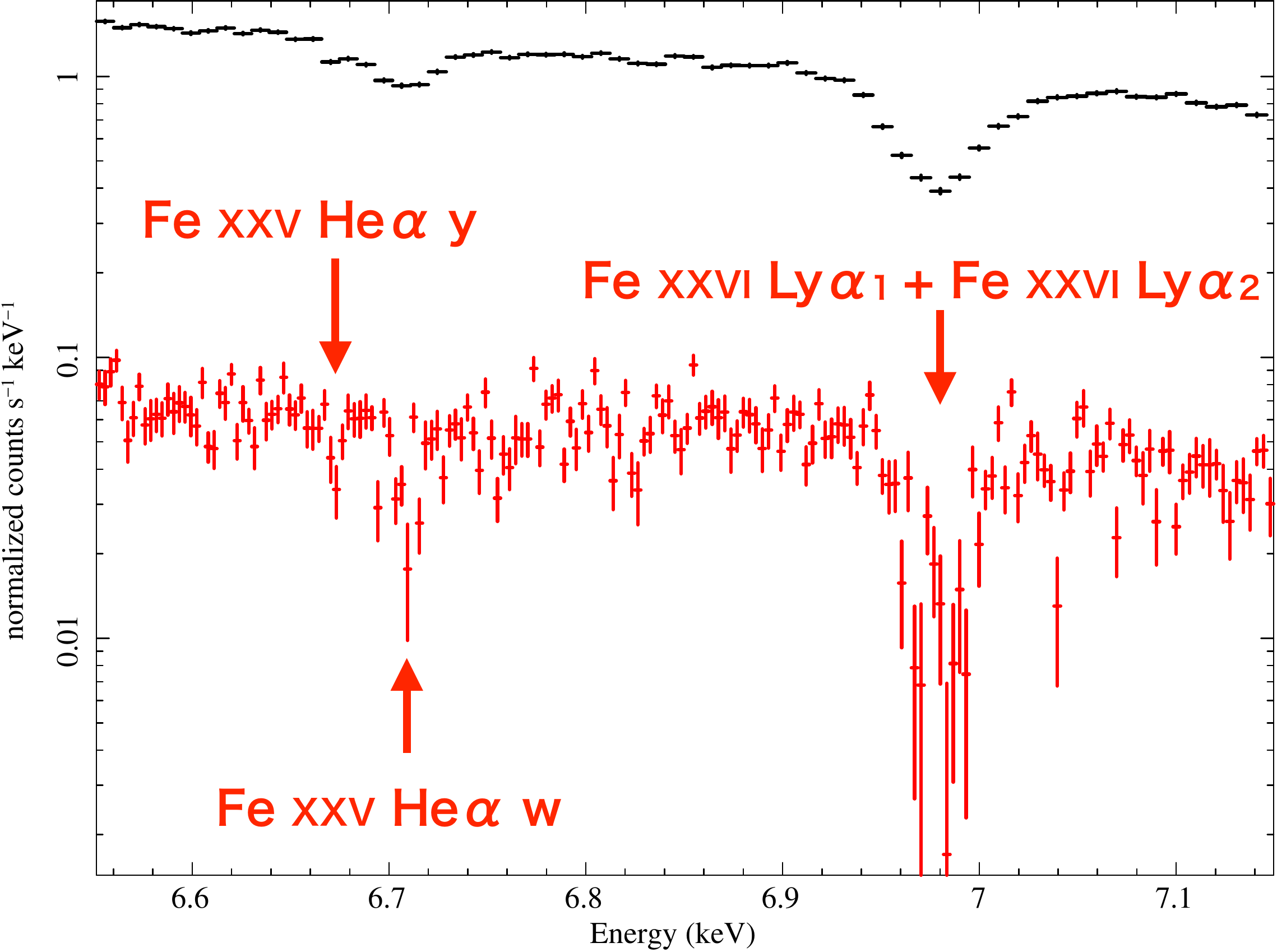}
\caption{HEG spectra of GX 13+1 from 1st order (black) and 3rd order
  (red). }
\label{fig:heg spectra}
  \end{figure}

GX13+1 was observed by the Chandra HighEnergy Transmission Grating (HETG) 4 times in two weeks
in 2010 (Table.~\ref{table:obsids}). The first order data are shown in \citet{D'Ai2014} and reveal multiple absorption
features from highly ionised elements (see also \citealt{Ueda2004} for similar features in an earlier observation). Higher order grating spectra give higher resolution, as demonstrated for the
black hole binaries by \citep{Miller2015}.  Here 
we show for the first time the third order Chandra data for GX13+1. 
We extract first- and third-order HEG spectra from these observations, 
using CIAO version 4.9 and corresponding calibration files.  We 
reprocess the event files with 
$"\mathrm{chandra\_repro}" $, and make response files using 
$"\mathrm{mktgres}"$ to make the redistribution and ancillary response files. 
We run  $"\mathrm{tgsplit}"$ to get the HEG $\pm 3$ spectra, and 
 run $"\mathrm{combine\_grating\_spectra}"$ to combine HEG plus and
minus orders for each observation to derive a single 1st order 
spectrum (black), and a single 3rd order spectrum (red) as shown in  
Fig.~\ref{fig:heg spectra}.  The 1st order spectra can resolve the
components of the He-like Fe triplet, with a clear dip to the low
energy side at the resonance line energy of 6.7~keV, but the H-like
Ly$\alpha_1$ and $\alpha_2$ are blended together. The higher resolution
of the 3rd order spectra is able to clearly separate the He-like
intercombination and resonance lines, and even the H-like Ly$\alpha_1$
and $\alpha_2$ \citep{Miller2015}.

\begin{table}
  \begin{tabular}{|c|c|c|c|} \hline
 OBSID & MODE & Date& Exposure (ks)\\ \hline 
 11815& TE-F & 24/07/2010 & 28 \\
 11816 & TE-F & 30/07/2010 &28 \\ 
 11814 & TE-F & 01/08/2010 & 28 \\
 11817 & TE-F & 03/08/2010 & 28 \\  \hline
   \end{tabular}
   \caption{List of the Chandra HETG observations}
  \label{table:obsids}
 \end{table}
 
\subsection{Model of GX 13+1}

\begin{figure}
\includegraphics[width=0.9\hsize]{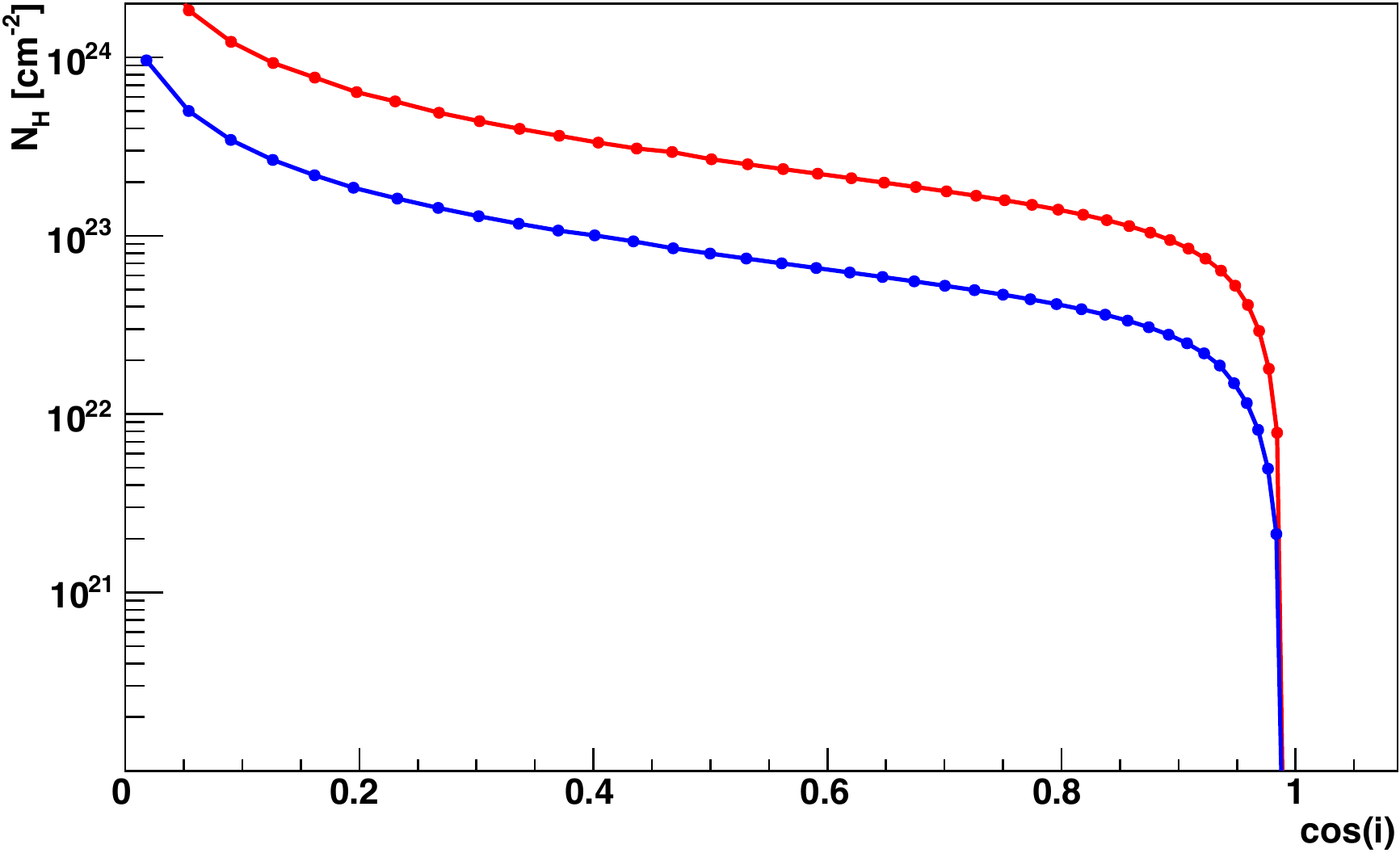}
\caption{The column density as a function of the cosine of the
  inclination angle for the diverging biconical wind calculated for the
  system parameters of GX13+1. The blue line shows the predictions for
  a purely thermal wind, while the red includes a very simple
  treatment of radiation pressure. The source has $L/L_{\mathrm{Edd}}\sim 0.5$,
  so the thermal wind can be launched from closer in due to the lower
  effective gravity. This effect has a large impact on the predicted column,
  so the details of how this radiation pressure correction affects the
  velocity and density structure will be important in determining the
  line profiles.}
\label{fig:radiation correction}
\end{figure}

We fit the contemporaneous RXTE spectrum (ObsID 95338-01-01-05) with a
model consisting of a disc, Comptonised boundary layer and its
reflection. The resulting inverse Compton temperature of the continuum
(disc plus Comptonisation) is $T_{\mathrm{IC}}\sim 1.2 \times
10^7~\mathrm{K}$, almost identical to the simulation (see also \citealt{D'Ai2014}). The luminosity is
$L=0.5L_{\mathrm{Edd}}$ \citep{DiazTrigo2014, D'Ai2014}, similar to the maximum
simulation value of $L=0.3L_{\mathrm{Edd}}$ in W96.  The simulation
also requires $ R_{\mathrm{out}}$, which can be calculated from the
orbital period and mass of binary stars. GX 13+1 has 24 day orbital
period, and the neutron star and companion have masses of $1.4
M_\odot$ and $5 M_\odot$ respectively \citep{Bandyopadhyay1999,
  Corbet2010}. This gives a binary separation $a=4.6\times10^{12}
\mathrm{cm}$, for a Roche-lobe radius $R_R/a=0.27$. The disc size is
then $R_{\mathrm{out}}= 10R_{\mathrm{IC}}$ assuming that
$R_{\mathrm{out}}=0.8R_R$ \citep{Shahbaz1998}, double the value
assumed in the simulations. D18 shows that this increase in disc size
makes the predicted column slightly larger, but the effect is fairly
small (Fig.~3: D18).  Fig.~\ref{fig:radiation correction} (blue line)
shows the predicted column density through the wind as a function of
inclination angle. This is very similar to the column predicted for
the fiducial simulations (Fig.~\ref{fig:column density})

However, D18 show that radiation pressure should make a rapidly
increasing contribution to the wind as $L/L_{\mathrm{Edd}}$ increases
from $0.3-0.7$.  The GX13+1 luminosity is midway between these two, so
radiation pressure should significantly lower the effective gravity,
meaning that the wind can be launched from smaller radii. We follow
D18 and estimate a radiation pressure correction to the launch radius
of $\bar{R}_{\mathrm{IC}} =
(1.0-0.5L_{\mathrm{Edd}}/0.71L_{\mathrm{Edd}})R_{\mathrm{IC}}=0.30R_{\mathrm{IC}}$
, hence $R_{\mathrm{out}}=33 \bar{R}_{\mathrm{IC}}$, dramatically
larger than assumed in the fiducial simulations. This correction
predicts a density which is $11$ times larger and column along any
sightline which is $ 3.3$ times larger assuming (as in D18) that the
velocity structure is unchanged (red line, Fig.~\ref{fig:radiation
  correction}).  This increase in $R_{\mathrm{out}}$ in terms of
$R_{\mathrm{IC}}$ means that more wind is produced (as in D18), so the
wind efficiency increases to 4.0 (from 2.3).

The column density goes
close to $10^{24}$~cm$^{-2}$ at high inclinations, so electron
scattering becomes important. This effect reduces the illuminating
ionising flux by $e^{-\tau_T}$ from the central source along the line
of sight to each wind element, and also increases the contribution of
diffuse and scattered emission from the wind to the ionising
continuum.  We include scattering, reducing the XSTAR illumination by
$e^{-\tau_{T}}$ along each line of sight, but do not include the
diffuse emission as the timescale to integrate over 
the entire wind at each point is prohibitive.

\begin{figure}
\includegraphics[width=0.9\hsize]{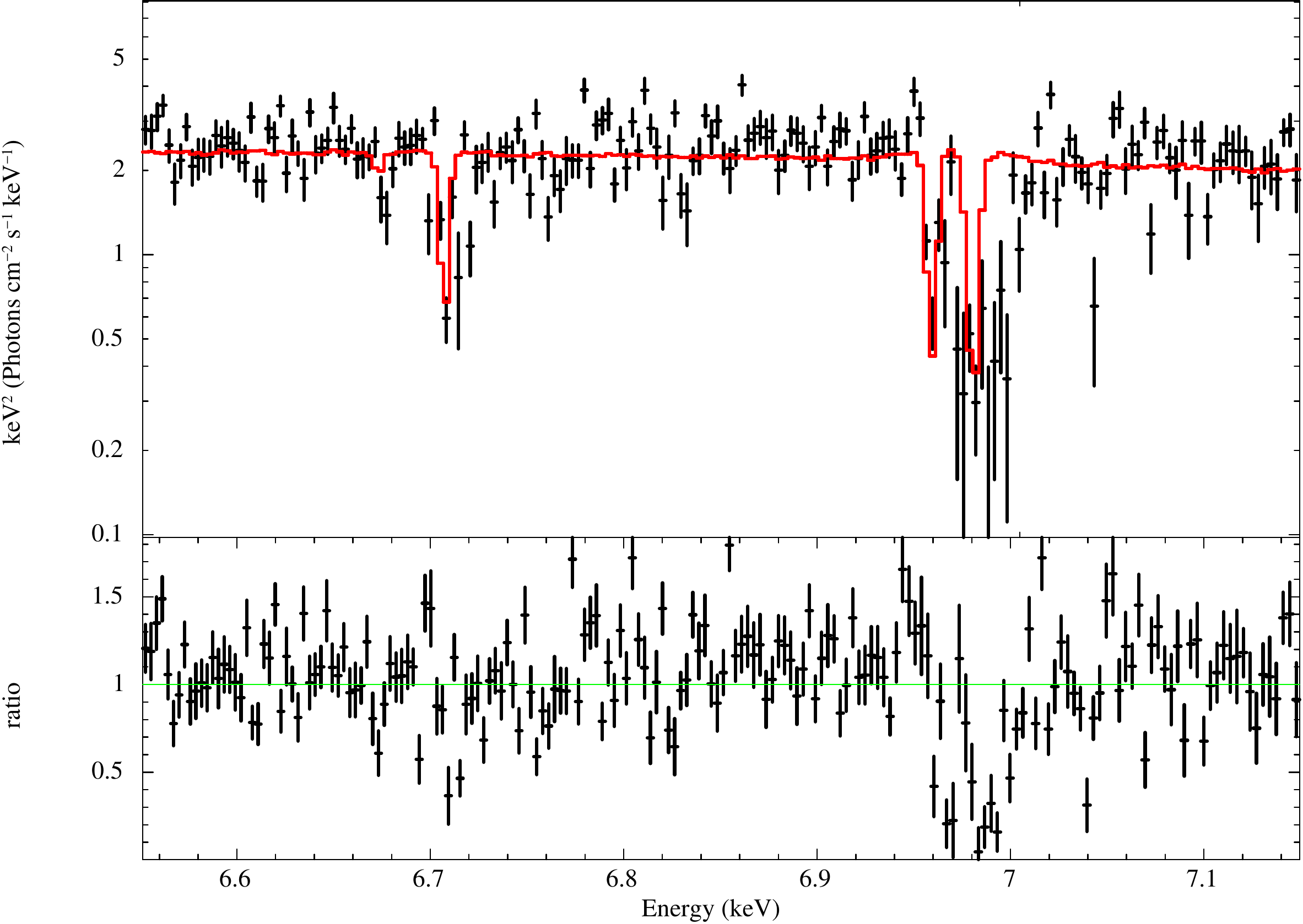}
\caption{The model (red) and HEG 3rd order spectrum (black). 
The best fit inclination angle is $i=80^\circ$. This gives roughly the
correct column of Fe {\scriptsize XXV} and {\scriptsize XXVI} at low velocity, but fails to match
the observed higher velocity blue wing to the absorption features. 
}
\label{fig:gx model1}
\end{figure}

\begin{figure}
\includegraphics[width=0.9\hsize]{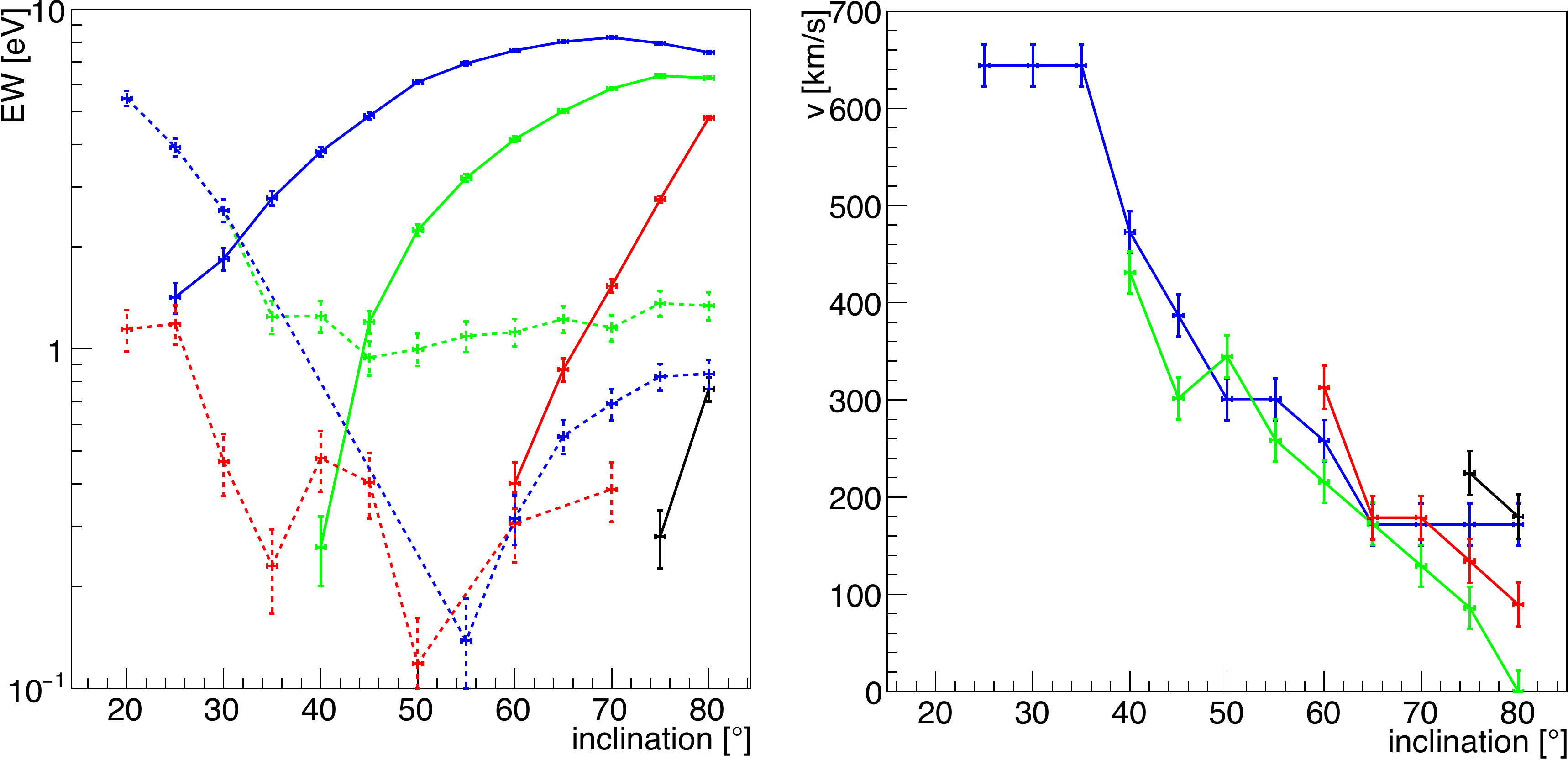}
\caption{As in Fig.~\ref{fig: ew d}, but with the system parameters 
of GX 13+1 and the simplest radiation pressure correction to make a
hybrid thermal/radiative wind.}
\label{fig:ew gx}
\end{figure}

We run MONACO on this wind structure to predict the detailed
absorption line profiles for comparison to the 3rd order HEG data.
Fig.~\ref{fig:gx model1} shows the result assuming an inclination
angle of $80^\circ$ \citep{DiazTrigo2012} which gives the best fit to
the data. This gives a fairly good match to the overall absorption,
except for the highest velocity material seen in the data.  Lower
inclination angles give higher blueshift, but lower absorption line
equivalent width, while higher inclination gives larger absorption
line but lower blueshift (see Fig.~\ref{fig:ew gx}).  Thus it is not
possible to completely reproduce the observed lines in GX13+1 with our
simple radiation pressure corrected thermal wind model.  This is not
surprising, as radiation pressure will almost certainly change the
velocity law by radiative acceleration as well as changing the launch
radius.  Full radiation hydrodynamic simulations are required to 
predict the resulting velocity and density structure. Nonetheless, our
result demonstrates for the first time that hybrid thermal-radiative
wind models can give a good overall match to the column and ionisation
state of the wind in GX13+1, and that current data can already give
constraints on the velocity and density structure of this material. 

\section{Discussion and Summary}

We construct a Monte-Carlo code to calculate detailed spectra from any
given density and velocity distribution of highly ionised material.
We use this to explore the absorption and emission lines of H and
He-like Fe for the mass loss rates predicted from thermal wind
models. We first use the radial streamline, constant velocity model of
D18 which is able to reproduce the column derived from the
hydrodynamic calculations of W96, but then extend this to a more
realistic disc-wind geometry with gas accelerating along diverging
streamlines, again reproducing the column from W96. The different
assumed velocity and density structures for the thermal wind mass loss
rates give different predictions for the overall ionisation state of
the material, the resulting EW of emission and absorption lines, and
their velocity shift. These show the potential of observations to test
the detailed structure of the wind.

We apply the biconical disc wind model to some of the best data on
winds from an LMXB. The neutron star GX13+1 shows strong and
persistent absorption features in Chandra first order HETG spectra
\citep{Ueda2004, D'Ai2014}, but here we show for the first time the
higher resolution third order data. We find that while the source is
fairly well matched to the parameters of the brightest fiducial
simulation in terms of $T_{\mathrm{IC}}$, the higher luminosity
\if{\bf\fi ($L/L_{\mathrm{Edd}}=0.5$ compared to $0.3$ for the simulation) makes
a significant impact on the predicted wind properties\if}\fi as it puts the
source firmly into the regime where radiation pressure driving should
become important.

We use the simple radiation pressure correction suggested by D18 and
calculate the line profiles from a hybrid thermal-radiative wind. The
additional radiation pressure driving means that the wind can be
launched from much closer to the central source, and has higher mass
loss rate.  This is the first detailed test of the absorption line
profiles predicted by physical wind models on any source other than
the singular wind seen in GRO J1655-40 \citep{Luketic2010}. Our
simulations quantitatively match many of the observed features except
for the highest velocity material.  This is not surprising, given the
simplistic assumptions about the effect of radiation pressure. In
future work we will use the wind velocity and density structure
determined from full radiation hydrodynamics simulations in order to
properly test the thermal-radiative wind models in
GX13+1. Nonetheless, our current simulations already show that the
thermal-radiative winds can potentially explain all of the wind
absorption features seen in GX13+1, so that there is very little
room for any additional magnetically driven winds in this source.



\section*{Acknowledgements}
We thank K. Hagino for help in setting up the MONACO simulations. CD acknowledges STFC funding under grant ST/L00075X/1
and a JSPS long term fellowship L16581.




\bibliographystyle{mnras}
\bibliography{library}

\appendix
\section{The rotation velocity for radial streamlines}
\label{sec:calculation model A}

Here we give details of how we calculate the rotation velocity of each
element of the wind for radial streamlines (Section 3).  We have a
linear radial grid, with 20 points from $R_{in}$ to $R_{out}$, so
spaced by $dR=(R_{out}-R_{in})/20$. The inner shell has midpoint
$R_0=R_{in}+dR/2$. We inject all the mass loss rate into this radial
shell, distributed as $(1-\cos\theta)$, on a linear grid of 20 points
in $\theta$. Each point on this inner shell is at a horizontal
distance of $R_0\sin\theta$ from the black hole. We assume the
material has the Keplarian velocity at this horizontal distance i.e. $
v_\phi(R_{0},\theta) = \sqrt{GM /(R_{0}sin \theta)} $.  Angular
momentum conservation along each stream line (of constant $\theta$ for
these radial streamlines) then gives $R_0\sin\theta
v_\phi(R_{0},\theta)=R\sin\theta v_\phi(R,\theta)$ so
$v_\phi(R,\theta)=(R/R_0) v_\phi(R_{0},\theta)$

\section{Density and velocity structure for the diverging streamlines}
\label{sec:calculation model B}


The diverging wind streamlines originate from the focal point which is
a distance $d$ below the black hole (see Fig 4).  The innermost edge
of the streamlines for the wind is at $\alpha_{min} =
\arctan(R_{in}/d)$, and the outer edge is at $\alpha_{max} =
\arctan(R_{out}/d)$. We make a linear grid so there are 40 angle
elements in the wind, separated by
$d\alpha=(\alpha_{max}-\alpha_{min})/40$ so that $\alpha_i=\alpha_0+i
d\alpha$ for $i=0\ldots 40$. We have $\alpha_0$ as a free parameter,
set by comparison to the results of W96 (see section 4) 

The maximum 'streamline' length below the disc is from $\alpha_{max}$, where
$D=\sqrt{d^2+R_{out}^2}$. We follow this for the same length above the
disc. This defines the outer radius of the simulation box which is 
$R_{max}=2\sqrt{d^2+R_{out}^2}$. We take the inner edge at
$R_{in}=0.1R_{IC}$.

\begin{figure}
\includegraphics[width=0.9\hsize]{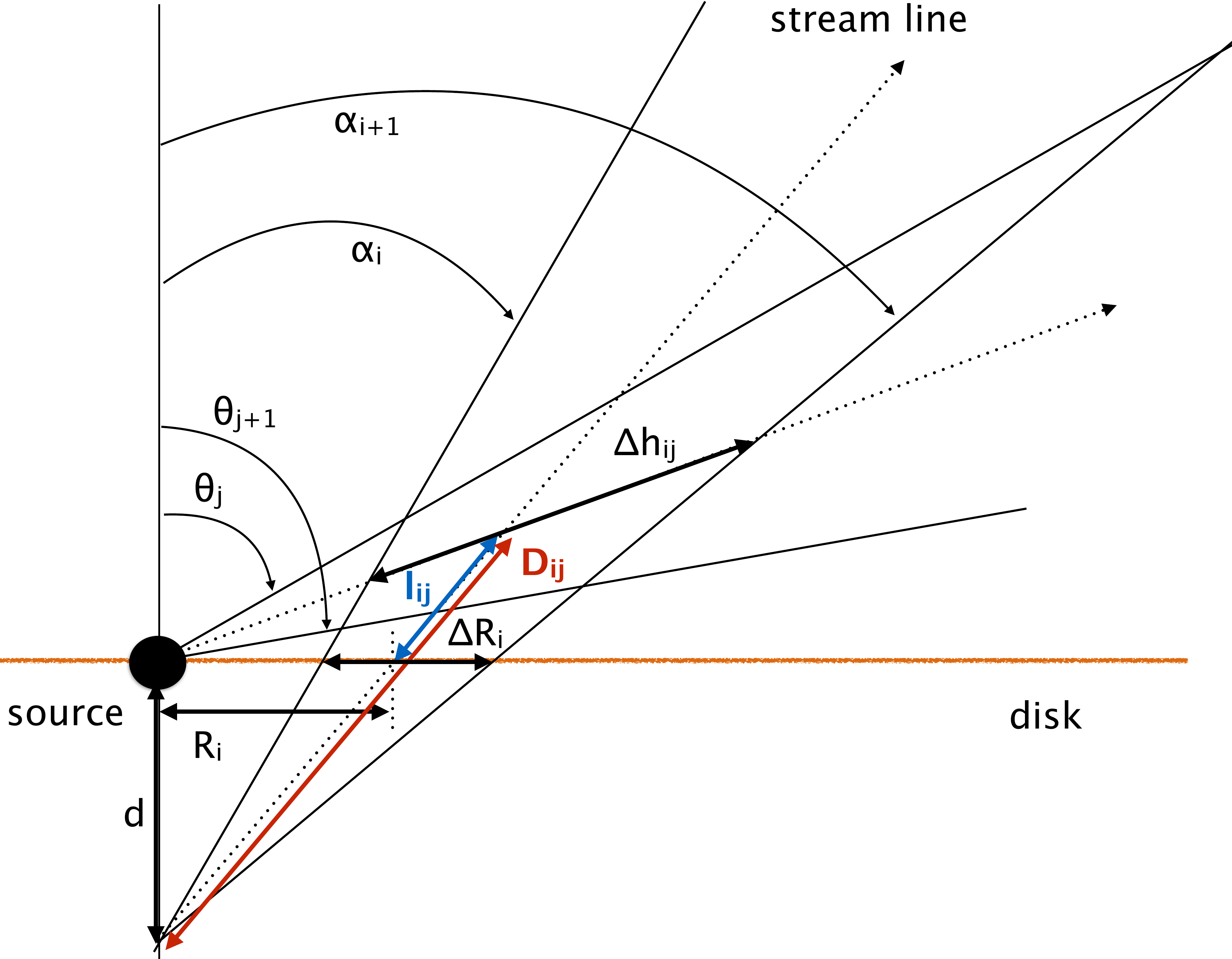}
\caption{Details of the model geometry for the diverging wind
  streamlines.}
\label{fig:shell B}
\end{figure}

We superpose a standard $\theta$ grid on this (measuring down from the
z-axis to radial lines from the centre: Fig.\ref{fig:shell B}). We set
$\theta_0$ to the point where the innermost streamline edge (at angle
$\alpha_0$) reaches $R_{max}$ from the origin, and take 41 angles from
this to $\pi/2$, giving $\theta_j (j=0,1..40)$.  We make shells using
the crossing points of these angles $\theta_j$ with the initial angles
$\alpha_i$ (Fig.\ref{fig:shell B}). We also define the midpoint angles 
$A_i=(\alpha_i + \alpha_{i+1})/2$ and 
$\Theta_j =\frac{1}{2}(\theta_j+\theta_{j+1})$ .  

The velocity along each stream line at a distance $l_{ij}$ from its
launch point on the disc at radius $R_i=d\tan A_i$ is
\begin{equation}
v_l (R_i,l_{ij})=f_v c_{ch}(R_i)\sqrt{\frac{l_{ij}}{R_i}}
\end{equation}
where
\begin{equation}
l_{ij}=D_{ij}-d/\cos A_i, D_{ij}=d\frac{\sin \Theta_j}{\sin(\Theta_j-A_i)}
\end{equation}
for a characteristic sound speed 
$c_{ch}(R_i)=\sqrt{\frac{kT_{ch}(R_i)}{\mu m_p}} $ defined from the
characteristic temperature
\begin{equation}  
T_{ch}(R_i)= \left( \frac{L}{L_{cr}} \right)^{2/3} (R_i/R_{IC} )^{-2/3} 
\end{equation}
where the critical luminosity, $L_{cr}$ is 
\begin{equation}
L_{cr}= \frac{1}{8} \left (\frac{m_e}{\mu m_p}\right )^{1/2} \left( \frac{m_e c^2}{kT_{IC}} \right)^{1/2} L_{Edd}
\end{equation} 
(see Done et al. 2018) and $f_v$ is free parameter which is determined
by comparing with the results of W96.

We calculate the density of each shell $n_{ij}$ assuming mass conservation along each streamline.
\begin{equation}
n_{ij}= \frac{\Delta \dot{M}_{wind}(R_i)}{m_I v_l(R_i, l_{ij}) 4 \pi D_{ij}^2 (\cos \alpha_i-\cos \alpha_{i+1})}
\label{eq:density}
\end{equation} 
where
\begin{equation}
\Delta \dot{M}_{wind}(R_i)= 2\pi \dot{m}(R_i) R_i \Delta R_i \times 2 =4\pi \dot{m}(R_i) R_i d(\tan \alpha_{i+1}-\tan\alpha_i)
\end{equation}
The total mass loss rate at a given luminosity $L$ 
is $\dot{M}_{wind} = \sum_i \Delta \dot{M}_{wind}(R_i) = 2.0\times 10^{19}  ~\mathrm{g/s} ~(L/L_{edd}=0.3), ~8.0\times 10^{18} ~\mathrm{g/s}  ~(L/L_{edd}=0.08),~ 2.1 \times 10^{18} ~\mathrm{g/s} ~(L/L_{edd}=0.01)$. 

Finally we, calculate the column density.
\begin{equation}
	N_H(\Theta_j)=\sum_i n_{ij}\Delta h_{ij}
\end{equation}
where
\begin{equation}
\Delta h_{ij}=d(\frac{\sin\alpha_{i+1}}{\sin(\Theta_j-\alpha_{i+1})}-\frac{\sin\alpha_i}{\sin(\Theta_j-\alpha_i)})
\end{equation}

We assume  Keplerian velocity on the disc plane ($\theta=\pi/2$ which
is at $j=40$) so that 
\begin{equation}
v_{\phi i,40}= \sqrt{\frac{GM}{D_{i,40}\sin A_i}}
\end{equation}
and assume the angular momentum conversation along stream line so that 
\begin{equation}
v_{\phi ij}=\frac{v_{\phi i,40}D_{i,40}\sin A_i}{D_{ij}\sin A_i}  =\frac{v_{\phi i,40} D_{i,40}}{D_{ij}}
\end{equation}

\bsp	
\label{lastpage}
\end{document}